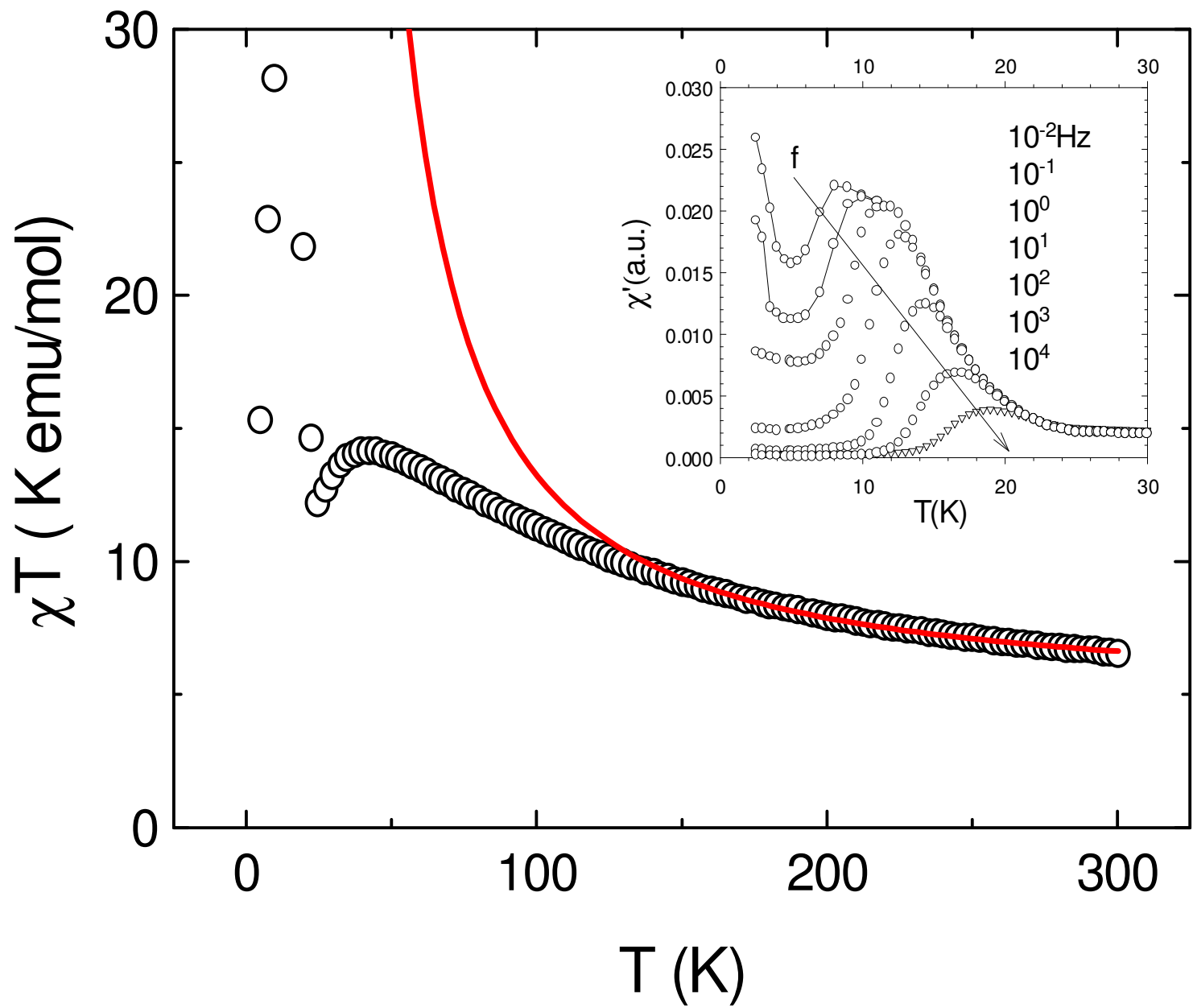

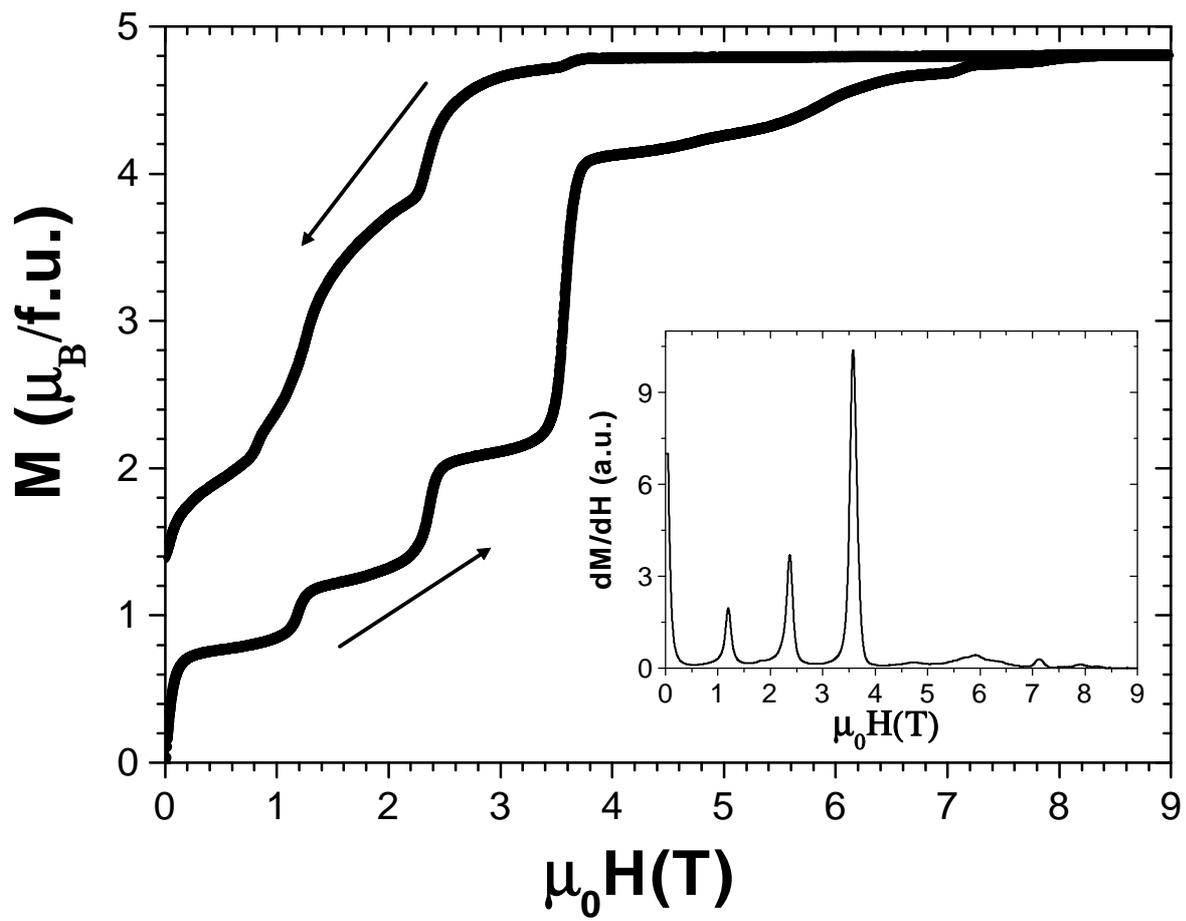

Figure 3

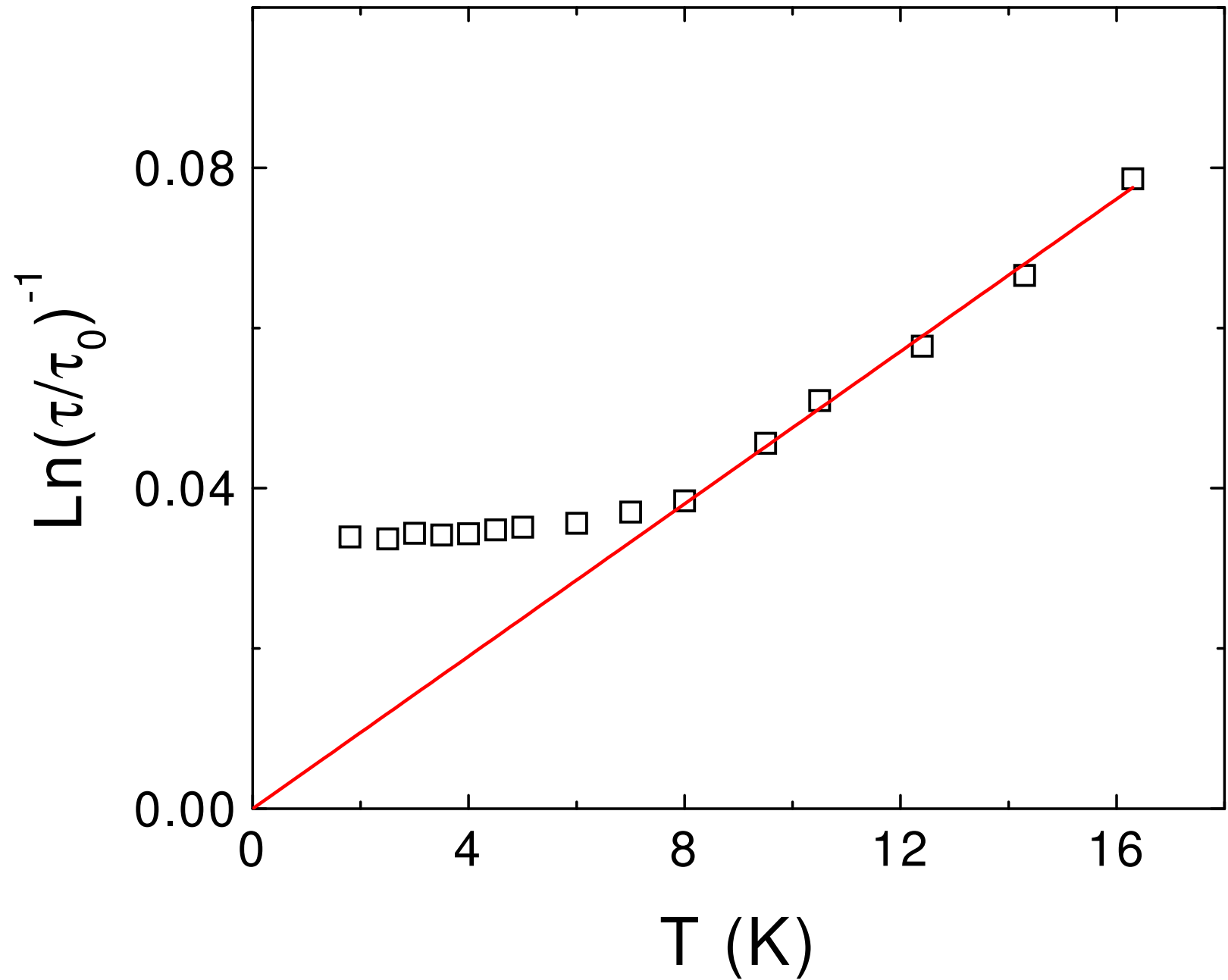

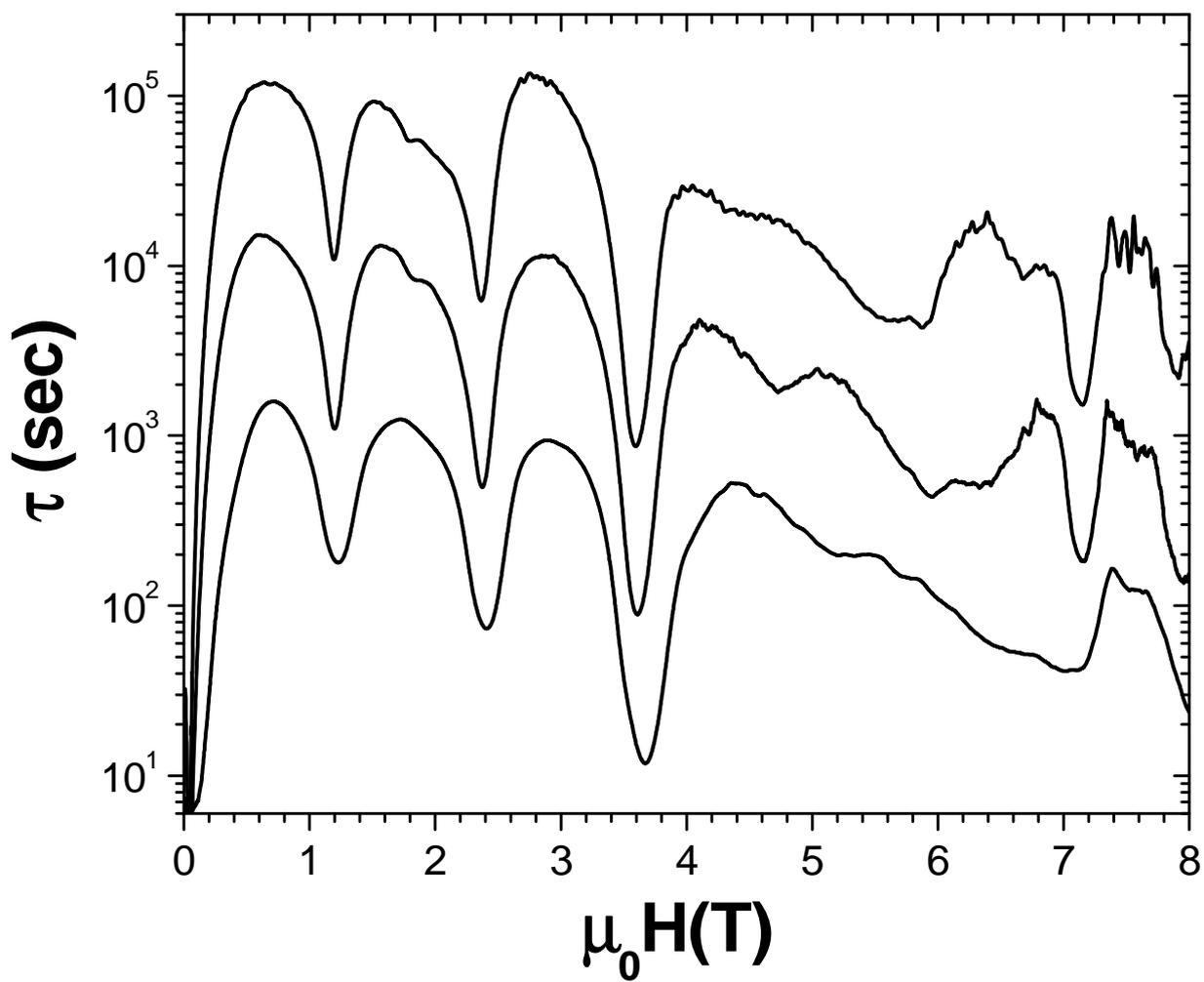

Fig5

# Quantum Tunneling of the Magnetization in the Ising Chain Compound $Ca_3Co_2O_6$

A. Maignan, V. Hardy and S. Hébert.
Laboratoire CRISMAT, UMR CNRS ISMRA 6508, 6 bd Maréchal Juin
14050 Caen Cedex, France

M. Drillon
IPCMS, UMR 7504, 23, rue du Loess, 67037 Strasbourg, France

M.R. Lees O. Petrenko and D. Mc K. Paul,
Department of Physics, University of Warwick, CV4 7 AL Coventry, United Kingdom

D. Khomskii
Physikalishes Institut, Universität zu Köln,
Zülpicher Str. 77, 50937 Köln, Germany

Abstract

The magnetic behavior of the $Ca_3Co_2O_6$ spin chain compound is characterized by a large Ising-like character of its ferromagnetic chains, set on triangular lattice, that are antiferromagnetically coupled. At low temperature, T < 7K, the 3D antiferromagnetic state evolves towards a spin frozen state. In this temperature range, magnetic field driven magnetization of single crystals (H//chains) exhibits stepped variations. The occurrence of these steps at regular intervals of the applied magnetic field, $H_{step}$=1.2T, is reminiscent of the quantum tunneling of the magnetization (QTM) of molecular based magnets. Magnetization relaxation experiments also strongly support the occurrence of this quantum phenomenon. This first observation of QTM in a magnetic oxide belonging to the large family of the $A_3BB'O_6$ compounds opens new opportunities to study a quantum effect in a very different class of materials from molecular magnets.

There are only few observations of quantum-mechanical effects at the macroscopic scale in magnetism among which the resonant tunneling of the magnetization has been recently the focus of much interest[1]. The most exciting results have been reported for the so-called single-molecule magnets [$Mn_{12}O_{12}(CH_3COO)_{16}(H_2O)_4$], Mn12Ac, and [$Fe_8O_2(OH)_{12}$ $(tacn)_6]^{8+}$, Fe8, which at very low temperature exhibit large hysteretic effects and regular steps in the magnetization, analyzed in terms of magnetic moment reversal by quantum tunneling[1-5]. In zero field, the macrospin of these molecules has two degenerate ground-states separated by a potential energy barrier, while in the magnetic field the symmetry is broken, and one spin-state is stabilized with respect to the other. As the field increases, a tunneling process through the energy barrier is made possible each time this state becomes resonant with a different excited quantum spin-state. As the temperature decreases from the spin blocking temperature, the magnetic relaxation evolves from an exponential decay, characteristic of a thermally activated process, to a square-root regime, that is assumed to be evidence of quantum tunneling. Basically, in these molecular compounds, the important ingredients are the high spin ground-state, that is well separated from the upper levels, and a large uniaxial magnetic anisotropy, which together are responsible for the spin-state splitting and the slow relaxation of the magnetization. Slow dynamics of the magnetization has also been predicted in chain compounds characterized by a strong uniaxial anisotropy[6]. It has been observed in quantum ferrimagnetic and ferromagnetic chains made of alternating $Co^{II}$ ions and radical moieties, with very slow relaxation of the magnetization and hysteresis effects[7].

In this paper, we focus on the magnetic properties of the chain compound $Ca_3Co_2O_6$, made of cobalt(III) ions occupying two distinct alternating sites in the chain, trigonal prism and octahedron[8], with the chains being set on a triangular frustrated lattice. Preliminary magnetic investigations[9] have emphasized the ferromagnetic character of the chains, with $Co^{III}$ spins being aligned along the chain axis, and a large amount of $Co^{III}$ ions in octahedral sites which are in the low spin (LS) state, S=0, as already observed for parent cobalt compounds such as $LaCoO_3$[10]. It is shown here that this unique situation in oxides, together with the large Ising-like character of the chain, favor the observation of a stepped variation of the magnetization which is discussed in terms of resonant tunneling.

The growth of $Ca_3Co_2O_6$ single crystals has been described elsewhere[11]. This compound consists of chains running along the c axis of the hexagonal unit cell which are made of alternating face-sharing octahedral and trigonal prismatic $CoO_6$ polyhedra, referred to as Co1 and Co2. Owing to the crystal field at Co1 sites, the spin state may be either high spin HS (S=2) or LS (S=0), while the smaller crystal field at Co2 sites promotes a HS state for all temperatures. These chains are arranged on a triangular lattice (Fig. 1) with an inter-chain distance of 5.24 Å, while the Co-Co separation within the chains is 2.59 Å[8].

Magnetic susceptibility measurements were performed in an ac field of 3 Oe and frequencies ranging from $10^{-2}$ to $10^4$ Hz, using extraction and SQUID magnetometers (MPMS and PPMS, Quantum Design). Magnetization measurements were carried out with a SQUID magnetometer (0-5 T) and a vibrating sample magnetometer (Oxford Instruments; 0-9 T). Relaxation experiments were recorded with the same equipment. It must be noted that a very good consistency was found between the magnetization data of the two kinds of crystals that exist for $Ca_3Co_2O_6$: needle-like shape[11] and rhombohedral-like shape[14].

The temperature dependent magnetic susceptibility recorded with the magnetic field H parallel to the chain axis is plotted in Fig.2 as $\chi T=f(T)$. In the high temperature $\chi T$ (T) exhibits an asymptotic behavior, described by $\chi = 4.89/(T-77.2)$ (emu.mol$^{-1}$), which agrees with one spin S=2 (g=2.55) per formula unit. On cooling the behavior is typical of a ferromagnetic low dimensional system. According to the very close energies of the low, intermediate and high spin states for trivalent cobalt sitting in an octahedron as reported for $LaCoO_3$, the value of the Curie constant is alternatively explained by considering a small amount of Co1 sites in HS state. Above 150 K, the behavior may be described by a S=2 Ising chain model[12], with g=2.55 and a magnetic exchange constant J=13.0 K (Fig. 2), which agrees with previous findings[13]. The strong divergence observed upon cooling may be related either to interchain interactions, or to some defects into the chains. Below about 40K, a first decay of the magnetic moment is observed, then at lower temperatures two characteristic anomalies appear. The first one evidenced at T = 25K corresponds to the 3D AF order revealed by neutron diffraction studies[9]. In this magnetic state, the ferromagnetic chains are antiferromagnetically coupled. A characteristic λ type anomaly is recorded at this temperature in specific-heat experiments[14], but the entropy jump indicates that the magnetic ordering is far from being complete. On further cooling, the $\chi T$ product exhibits a strong increase down to about 14K, then a drop to zero. This second maximum (not visible on the figure) is strongly dependent on the timescale of the experiments, as evidenced in Fig.2 (inset) from the in-phase (χ') signal, which reflect a slow relaxation of the spins according to a superparamagnetic behavior[11]. Furthermore, at the lowest temperatures, typically below 7K ( in fact this temperature depends closely on the timescale of the experiment), there is a significant irreversibility in the magnetization curves, indicative of a spin frozen state which is also confirmed by specific heat measurements[14]. As shown in ref. 9, although the integrated intensity of the antiferromagnetic reflections increase below $T_N$ = 25K as expected, below 15K a decrease of these integrated intensities was found. This unusual T dependence for an antiferromagnet can be ascribed to the onset of the spin frozen state. Clearly, a loss of magnetic coherence in the chains is induced by decreasing T which promotes the creation of finite spin units in the chains. At low temperature, for instance at 2K, the M(H) curve exhibits a remarkable stepwise variation (Fig.3). On closer examination, despite the apparent complexity, these steps are shown to occur at regular intervals of the applied field, $H_{step}$=1.2T, while their height depends on temperature. This is clearly demonstrated from the dM/dH versus H plot (inset of Fig.3) for T=2 K. Although these periodic steps are more difficult to observe for H > 3.6T, there are four additional steps separated by 1.2T, which correspond to about 15% of the total magnetization. Notice that a large plateau is observed at 4.1$\mu_B$/f.u., but the actual saturation (4.8$\mu_B$/f.u.) is in fact obtained for H > 8T. It should also be pointed out that at 10K a unique step of height M=1.6 $\mu_B$/f.u. at H=3.6T is seen, which corresponds precisely to one third of the saturation magnetization, according to an AF coupling between chains. It must be emphasized that for an applied magnetic field perpendicular to the chain axis, the M(H) curves exhibit a linear variation with very low magnetization values, which confirm the very strong Ising-like anisotropy of the chains.

In order to compare these results with those of systems exhibiting quantum tunneling phenomena, we have probed the spin dynamics of $Ca_3Co_2O_6$ at different temperatures. The values of the relaxation time $\tau$ have been extracted, in the range 6-16K, from the position of the maxima of $\chi$" (for f =$10^{-2}$ Hz to $10^4$ Hz), and by the relaxation of the magnetization at lower temperatures. After application of a 5T magnetic field, the magnetization measured at H = 0 exhibits a strong decay with time which was fitted, over large timescales, by a stretched exponential M = $M_0$ + $M_1$exp[-((t+$t_0$)/$\tau$)$^{0.5}$], usually used for describing the behavior of molecule-based magnets at low temperature. It should be noted that $M_0$ is 50% of the remanent magnetization ($M_0$+$M_1$) at 1.8 K and 19 % at 5 K. The variation of the relaxation time is plotted as [ln($\tau/\tau_0$)]$^{-1}$ vs T in Fig.4, where $\tau_0$=5 $10^{-10}$s is found from an extrapolation of the data to high temperature. Above 7 K, the data are linear with T, in agreement with the thermally activated law $\tau = \tau_0 \exp(\Delta/kT)$, with $\Delta$ ~ 210 K. In the very low temperature range, the relaxation time shows an asymptotic behavior towards a constant value $\tau \sim 10^3$ s, a behavior consistent with a quantum tunneling mechanism. It must also be emphasized that the change of relaxation mechanism, at a blocking temperature of about 7K, is in good agreement with the spin frozen state revealed by specific heat measurement below that temperature. The relaxation of the magnetization at nonzero field may further be described by the expression $M = M_{eq}[1-\exp(-t/\tau)]$, where $M_{eq} = M_s \tanh(g\mu_B SH/kT)$, $M_{eq}$ and $M_s$ being the equilibrium and saturation magnetization. We then deduce the relaxation rate $\tau^{-1} = (dM/dH)(dH/dt)/[M_{eq}-M(H)]$ where $dH/dt$ is the sweeping rate of the field. Using the experimental value of the saturated magnetization corresponding to gS=4.8, one obtains the field dependence of the relaxation time plotted in Fig. 5 for T=2K. Striking features are present including (i) steep variations of the relaxation time with minima for H=1.2n (T) with n=1-6, (ii) the regular increase of the peak magnitude for n=1 to 3, (iii) the large peak at 7.2T, which is precisely twice the value for the major peak (iv) and finally a tiny bump at 1.8T which is half the value of the field at which the largest peak appears. Another important feature is that changes in the sweeping rate of the field (from 0.01 to 1 T.min$^{-1}$) do not modify the positions of the peaks, as expected for a quantum tunneling process. Furthermore, the small peak at H=1.8T disappears for the fastest sweeping rate.

Basically, a large Ising-like anisotropy together with a finite value of the spin are needed to promote tunneling process. In the model used to describe single molecule magnets, the simplest spin Hamiltonian is: $H = -DS_z^2 - g\mu_B SH$, where D stands for the anisotropy constant. According to this expression, spin crossing is achieved for regular intervals of the field, $H_{step} = -D/g.\mu_B$, which correspond to minima in the relaxation time of the magnetization. For the compound under consideration, the regularly spaced dips of the relaxation rate for $H_n = 1.2n$ (T) agree with such a model, but unlike single molecule magnets, the interaction between spin units cannot be neglected. The influence of these (interchain) antiferromagnetic interactions, characterized by a coupling constant J' of one chain with its z neighbors, is clear in the M(H) plot at low temperature which shows, upon decreasing field after saturation, the first steps appearing before having change the sign of the applied field. Writing for an interacting system the internal field as $H = H_{ext} + zJ'<S>$, we can deduce that $H_{ext} = -zJ'<S>$ is nothing more than the value of the external field compensating the spin-spin interactions.

An analysis of the above findings may be proposed by considering the formation of intrachain finite spin units into the so-called spin frozen state. Let us start from the extreme situation where the ferromagnetic coupling between HS Co2 through a LS Co1 is suppressed.

In this case, we are dealing with a collection of isolated single S = 2 spins. The presence of some S=2 trivalent cobalt sitting on octahedral sites can produce however trimers. In the same ways, dimers may be created by inverting the spin-states of two adjacent Co1 and Co2. If one assumes that for single Co the anisotropy constant is $D/g\mu_B = 3.6T$ (D is used for the single spin S=2), the Hamiltonian for m such cobalt spins will be : $H = -\Sigma(D(s_z)^2 + g\mu_B H s_z)$, the sum being over the m Co spins. If $s_z$ and $\mathbf{S}_z$ are the z projections of the individual $Co^{3+}$ spin and the cluster of m spins, respectively, the Hamiltonian can be rewritten : $H = -Dms_z^2 - g\mu\, m s_z$. Then, it turns out that $H = -(D/m)\mathbf{S}_z^2 - g\mu_B H\mathbf{S}_z$ where $\mathbf{S}_z = ms_z$, i.e. the anisotropy constant for the m spin unit is $\mathbf{D}=D/m$. Thus, if the anisotropy for a single HS Co is $D/g\mu_B = 3.6T$ (namely 6.03 K for g=2.55), the large peaks observed at 3.6 and 7.2 T can be explained by the quantum-tunneling relaxation process of single spins S=2, leading to characteristic fields H(T) = 3.6n. The other peaks can then be explained as originating from trimers [H(T) = 1.2n] and dimers [H(T) = 1.8n]. The fact that 1.2n jumps are the most prominent is consistent with the fact that HS octahedral $Co^{3+}$ ion (leading to the formation of trimers) is more probable than LS Co in prismatic site (producing dimers). It is to be noted that cross relaxation process in AF exchange-coupled units, or exchange-bias tunneling, may also provide the same features, as recently reported by Wernsdorfer et al[5]. For instance, AF dimeric units involving spins of different chains may also be at the origin of the transition at 1.8 T.

In conclusion, $Ca_3Co_2O_6$, that shows magnetization steps at fixed magnetic field intervals, whatever the timescale of the experiment, is likely to be the first example of a transition metal oxide that exhibits resonant magnetization tunneling. The tunneling mechanism probably involves spin units with S=2, S = 4 and S=6. This observation for a member of the series $A_3BB'O_6$[15], where B and B' are trivalent metal ions, opens new perspectives for the study of the resonant tunneling of magnetization.

**Acknowledgment :** The authors thanks P. Panissod (IPCMS) for fruitful discussions

**Figure Captions :**

Fig. 1: Schematic drawings of the $Ca_3Co_2O_6$ structure. The dark and light polyhedra are for the $CoO_6$ trigonal prisms and $CoO_6$ octahedra, respectively. The calcium cations are the small circles located in between the chains. (a) Perspective view showing the $[ABO_6]_\infty$ chains running along the hexagonal c-axis. (b) Projection along the c-axis. The solid lines show the triangular array of the chains in the ab plane.

Fig. 2: Temperature dependence of the $\chi T$ product for $Ca_3Co_2O_6$ crystals with the magnetic field applied along the chain axis. The curve corresponds to the $S = 2$ Ising chain fitting at high temperature (see text). Inset : Temperature dependence of the ac magnetic susceptibility ($\chi'$, real part) collected at different frequencies labelled in the graph.

Fig. 3: Field dependence of the magnetization of $Ca_3Co_2O_6$ crystals at 2K with the magnetic field applied along the chain axis. H sweeping rate : $0.1 T.min^{-1}$. The arrows indicate the increasing and decreasing field branches. The magnetization data are collected after a zero field cooling process down to 2K. Inset : field dependence of the corresponding derivative dM/dH.

Fig. 4: Temperature dependence of $Ln(\tau/\tau_0)^{-1}$ where the values of the relaxation time $\tau$ are extracted from $\chi'(T)_f$ and magnetization relaxation experiments. The linear fitting of the high T part of the curve corresponds to the thermally activated law discussed in the text.

Fig. 5: Magnetic field dependence of the relaxation time $\tau$ derived from the M(H) curves (see text) with different sweeping rates of the field (from top to bottom $0.01 T.min^{-1}$, $0.1 T.min^{-1}$ and $1 T.min^{-1}$).